\documentclass[]{spie}  

\usepackage[utf8]{inputenc}
\usepackage{graphicx}
\usepackage{ragged2e}
\usepackage{wrapfig} 
\usepackage{epstopdf}
\DeclareGraphicsExtensions{.pdf,.png,.jpg}
\usepackage{multirow}
\usepackage{varioref}
\usepackage{multicol}
\usepackage{float}
\usepackage{rotate}
\usepackage{subcaption} 
\usepackage{amsmath,amsfonts,amssymb}
\usepackage{graphicx}
\usepackage[colorlinks=true, allcolors=blue]{hyperref}

\title{Revival of an abandoned telescope: the Boller and Chivens Bochum 0.61-metre telescope of Universidad de Valparaiso}

\author[a]{Sebastián Zúñiga-Fernández}
\author[a]{Maja Vu\v {c}kovi\'{c}}
\author[a]{Nikolaus Vogt}
\author[a]{Omar Cuevas}
\author[b]{Yerko Chacon}
\affil[a]{Instituto de Física y Astronomía, Universidad de Valparaíso, Avenida Gran Bretaña 1111, Casilla 5030, Valparaíso, Chile}
\affil[b]{Observatorio Pocuro, Avenida Pedro Aguirre Cerda s/n sector Pocuro, Calle Larga, Chile}

\authorinfo{Further author information:\\ S.Z.F.: E-mail: sebastian.zuniga@postgrado.uv.cl \\ M.V.: E-mail: maja.vuckovic@uv.cl}

\begin{document} 
\maketitle
\renewcommand{\arraystretch}{2}%
\begin{abstract}
In 2015 the Institute of Physics and Astronomy of the Universidad de Valparaíso in Chile received as a donation the Bochum 0.61-meter telescope. Here we preset the ongoing project to convert this senior member of La Silla Observatory to modern standards aiming at performing state-of-art science, as well as teaching and outreach. Firstly, the site characterization was performed in order to verify the observing conditions. The preliminary results were auspicious in relation to the nights available for observation. In early 2016 began the transfer work form La Silla Observatory to the new site of operations. The actual status of the telescope was analyzed and an upgrade plan was proposed to make it usable remotely using a web-based telescope control system developed in Chile by ObsTech SpA. Future upgrade and scientific collaboration will be discussed based on the site characterization and technical studies regarding the potential for new instrumentation.
\end{abstract}

\keywords{Bochum 0.61-meter, Telescope, Telescope Control System, Site characterization, Observatorio Pocuro, Pointing, Facility upgrades}

\section{INTRODUCTION}
\label{sec:intro}  
The Bochum 0.61-meter telescope, installed on La Silla in 1968, was the first national telescope in ESO's history through a trilateral agreement between ESO, the Deutsche Forschungsgemeinschaft (German Research Foundation) and the University of Bochum. In the early 70s this telescope was one of the so-called ``beacon telescopes'' in the mountain, given its remarkable photo-spectrograph and also is remembered to be one of the first telescope to detect unexpected evolution phenomena from supernovae SN 1987A, still referred to as ``Bochum event'' in scientific literature \cite{hanuschik_1988}.    

The new professional telescope now of the Universidad de Valparaíso (including its original mounting and copula) was driven by a dream of Dr. Nikolaus Vogt, Professor and Researcher of the Institute of Physics and Astronomy (IFA) of the Universidad de Valparaíso (UV) in Chile, who made the arrangements with the University of Bochum to make the donation of this astronomical professional instrument, that was not being used anymore on La Silla, to the UV. 

\begin{figure}[H]
	\centering
	\includegraphics[width=.6\textwidth]{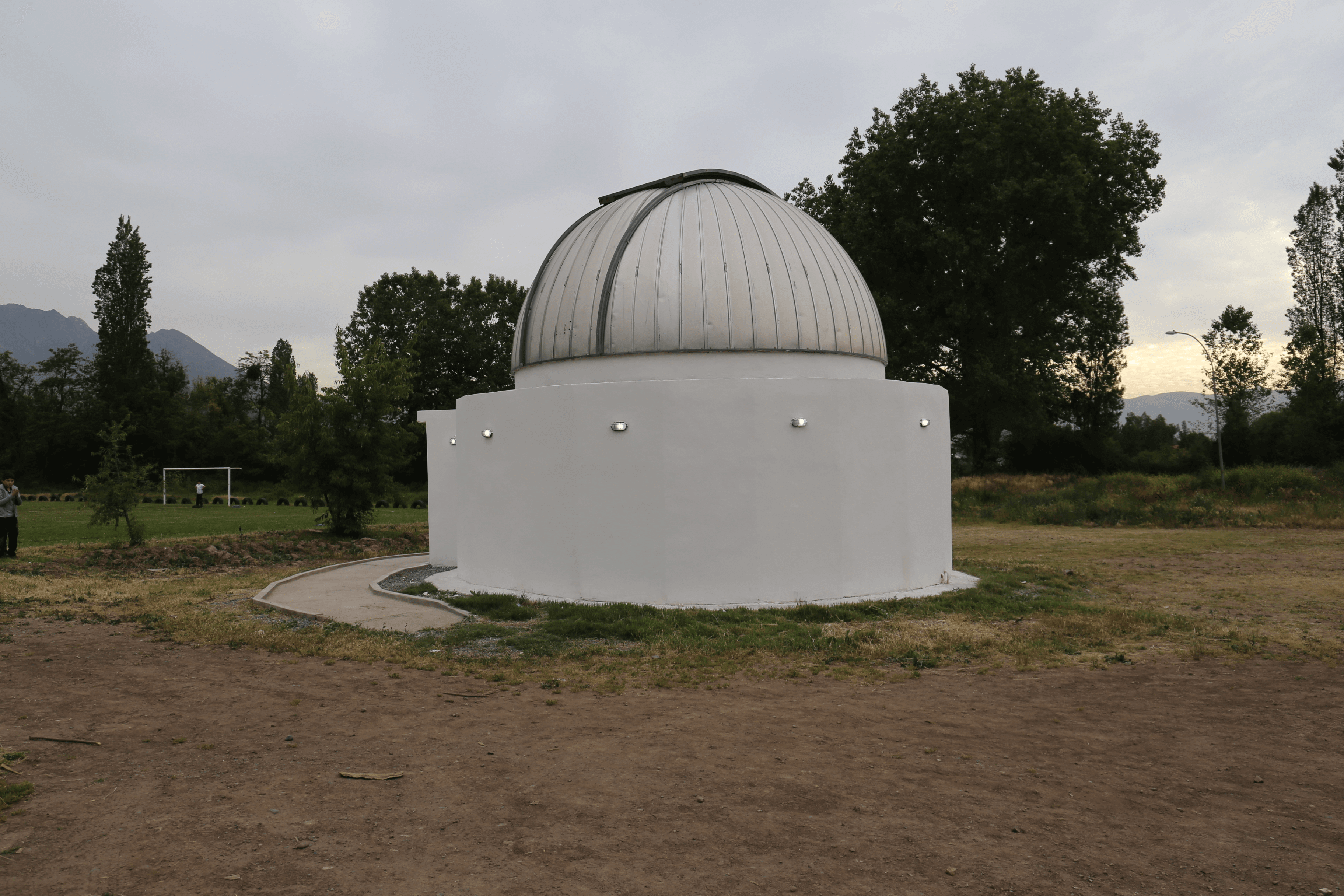}
	\caption{The Bochum 0.61-meter telescope at Observatorio Pocuro.}
	\label{fig:pocuro_telescope}
\end{figure}

Studies were carried out to find the observationally most optimal cite in the Valparaíso region. Finally, the ideal location was found to be in Pocuro at Calle Larga town (around 2 h from Valparaíso city). At Calle Larga  already exist active amateur astronomer community called \textit{Asociación Astronómica de Aconcagua} who founded the Observatorio Pocuro (including a set of small scale telescopes). After a collaboration agreement between the UV and Calle Larga authorities, the Bochum 0.6-meter telescope was placed in its new home at Observatorio Pocuro (see Fig. \ref{fig:pocuro_telescope}).

In early 2016 the transfer preparation work began  to transport the Bochum 0.6-meter telescope from La Silla Observatory to the new site of operations in the Observatorio Pocuro at Calle Larga. The transfer process involved management and logistic around safely transportation of the cupola and telescope. A new building was constructed at the Observatorio Pocuro in order to set up the cupola and the telescope properly.

Our goal is to modernize the Bochum telescope by upgrading the Telescope Control System (TCS) in order to replace the DOS-based system currently installed, along with the old power control electronics and encoders. The actual status of the telescope was analyzed and an upgrade plan was proposed to make it usable remotely using a web-based telescope control system developed in Chile by ObsTech SpA \cite{1mSilla}. 

The aim of the renovated Bochum telescope is multipurpose. Our project explores the huge potential of the small previously abandoned telescope for teaching, science and outreach purposes. This potential will be discussed and future plan will be presented.

\begin{table}[h!]
  \begin{center}
    \label{tab:table1}
    \begin{tabular}{|l|c|} 
    \hline
      \textbf{Enclosure} & Classical dome \\
      \hline
      \textbf{Type} & Photometric telescope \\
      \hline
      \textbf{Optical design} & Cassegrain reflector \\
      \hline
      \textbf{Diameter Primary M1} & 0.61 m \\
      \hline
      \textbf{Material Primary M1} & Low expansion Silica \\
      \hline
      \textbf{Diameter Secondary M2} & 0.15 m \\
      \hline
      \textbf{Material Secondary M2} & Low expansion Silica \\
      \hline
      \textbf{Mount} &  Equatorial cross-axis mount \\
      \hline
      \textbf{First Light date} & 7 September 1968 \\
      \hline
    \end{tabular}
    \caption{Bochum 0.61-meter telescope summary.}
  \end{center}
\end{table}

\section{Site information and future characterization}
The search for the place to locate the telescope, as in the large observatories, was not only based on the quality of the skies but also on the logistic conditions in order to be able to maintain and operate the telescope properly thoughout the year. In that respect the Observatorio Pocuro at Calle Larga turned out to be the best place in the Valparaíso region as it fulfills both of the above requirements.

Preliminary site characterization was developed in order to verify the observation conditions. The results were auspicious, revealing an average of 270 clear nights per year during the past 30 years. New studies are underway to measure specific parameters such as seeing, wind speed and humidity, including in the near future a meteorological and dim station.

\section{Beginning the Journey: From La Silla to Observatorio Pocuro}
\label{sec:transportation}
After receiving the telescope as a donation, the next challenge was to transport it to its new location, the Observatorio Pocuro, which is located approximately 600 km to the south of La Silla Observatory.

\subsection{Transportation from La Silla}
The first step before moving the telescope was to dismount it, in this way a safer and controlled transfer is guaranteed. It should be noted that the primary mirror (M1) and the secondary mirror (M2) were left at La Silla Observatory to be aluminized and were moved later.

\begin{figure}[H]
	\centering
	\includegraphics[width=.44\textwidth]{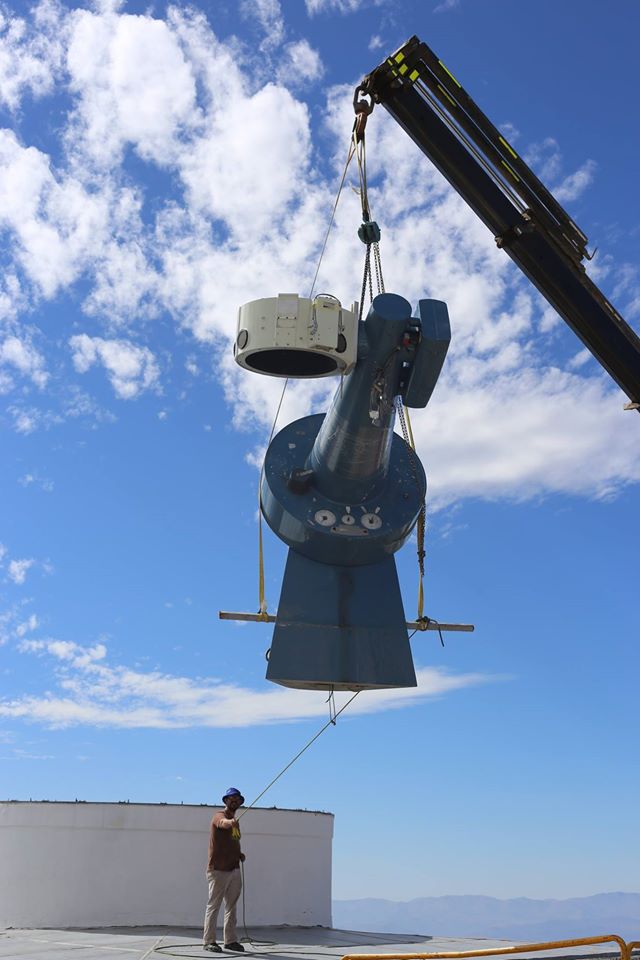}
	\caption{Crane truck lifting the equatorial mount at La Silla observatory.}
	\label{fig:telescopio_grua}
\end{figure}

The parts of the telescope were removed one by one in the following order: inner door of the dome, counterweights, mirrors M1 and M2, the secondary ``spider'', optical tube and finder, the ash-dome and finally the equatorial mount. For these procedures it was necessary to take a truck crane that would fulfill both height and load capacity requirements (see Fig. \ref{fig:telescopio_grua}).

\begin{figure}[H]
	\centering
	\includegraphics[width=.55\textwidth]{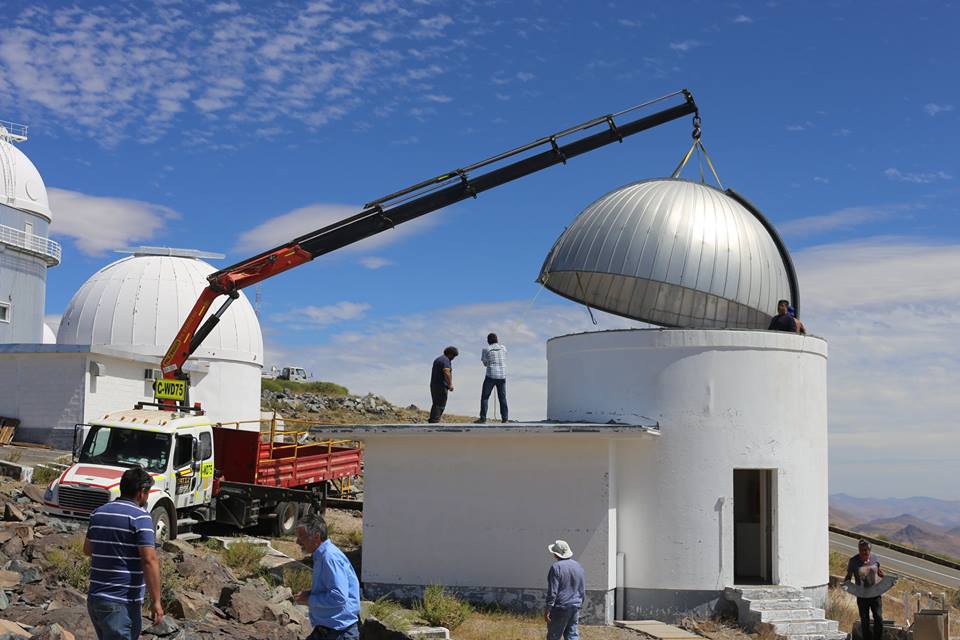}
	\caption{Lifting the dome of the Bochum telescope at the La Silla Observatory.}
	\label{fig:cupula_grua}
\end{figure}

For safety transportation, the dome was also dismounted including the main engines responsible for opening and rotating the dome (Fig. \ref{fig:cupula_grua}). First the main engine, that is supported in the upper part of the dome and that generates the opening of the hatch, was disassembled and the hatch was completely dismounted. Finally, the rotation motors of the dome were removed.

\subsection{Arrival at the Observatorio Pocuro}
It was necessary to mount a concrete central pillar aligned with the meridian of the site to support the equatorial mount. A new building was constructed according to the structural requirement to mount the original dome of the telescope. It is worth mentioning that the building has a 2 meter larger surface than the original building in La Silla in order to meet not only science needs, but also both teaching and outreach purposes (see Fig. \ref{fig:pocuro_telescope}).

\begin{figure}[H]
	\centering
	\includegraphics[width=.33\textwidth]{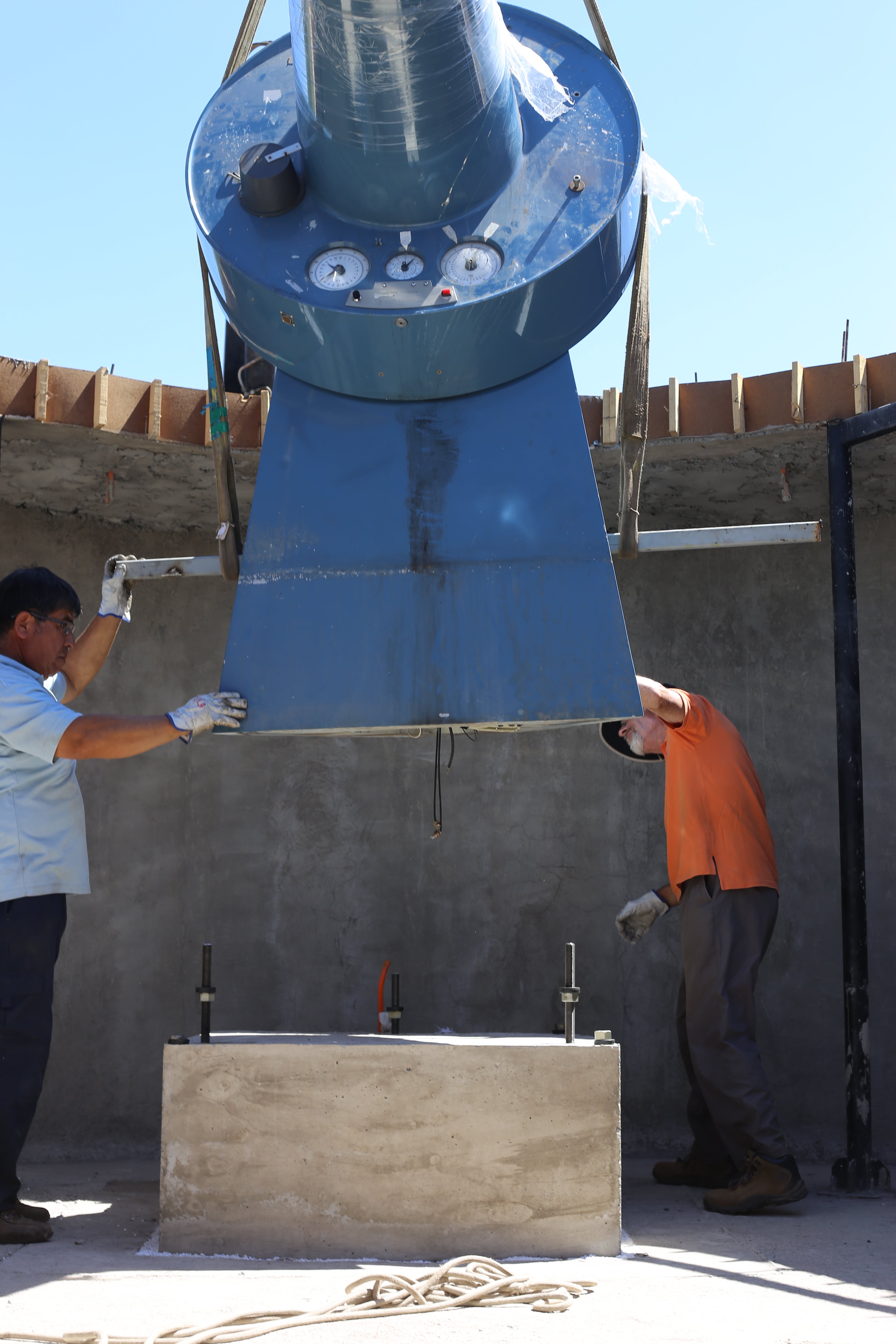}
	\caption{Lowering the equatorial mount onto the concrete pillar inside the dome building.}
	\label{fig:pillar_mount}
\end{figure}

The telescope was mounted on the concrete pillar and attached to three bolts with nut and lock nut (see Fig. \ref{fig:pillar_mount}). After the mount was installed, we proceeded to assemble the frame and install its counterweights. Finally, optical tube and finder were installed.

The dome was assembled piece by piece and lubricating every piece to ensure a smooth 360$^{\circ}$ rotation. Once the dome was assembled, it was lifted and installed in the new building. The building has three-phase current for operation. It should be noted that the rotation motors of the dome had to be repaired. The rotation of the dome has been one of the most complicated factors due to the subtle differences in the final terminations of the dome supporting ring.

\begin{figure}[H]
	\centering
	\includegraphics[width=.56\textwidth]{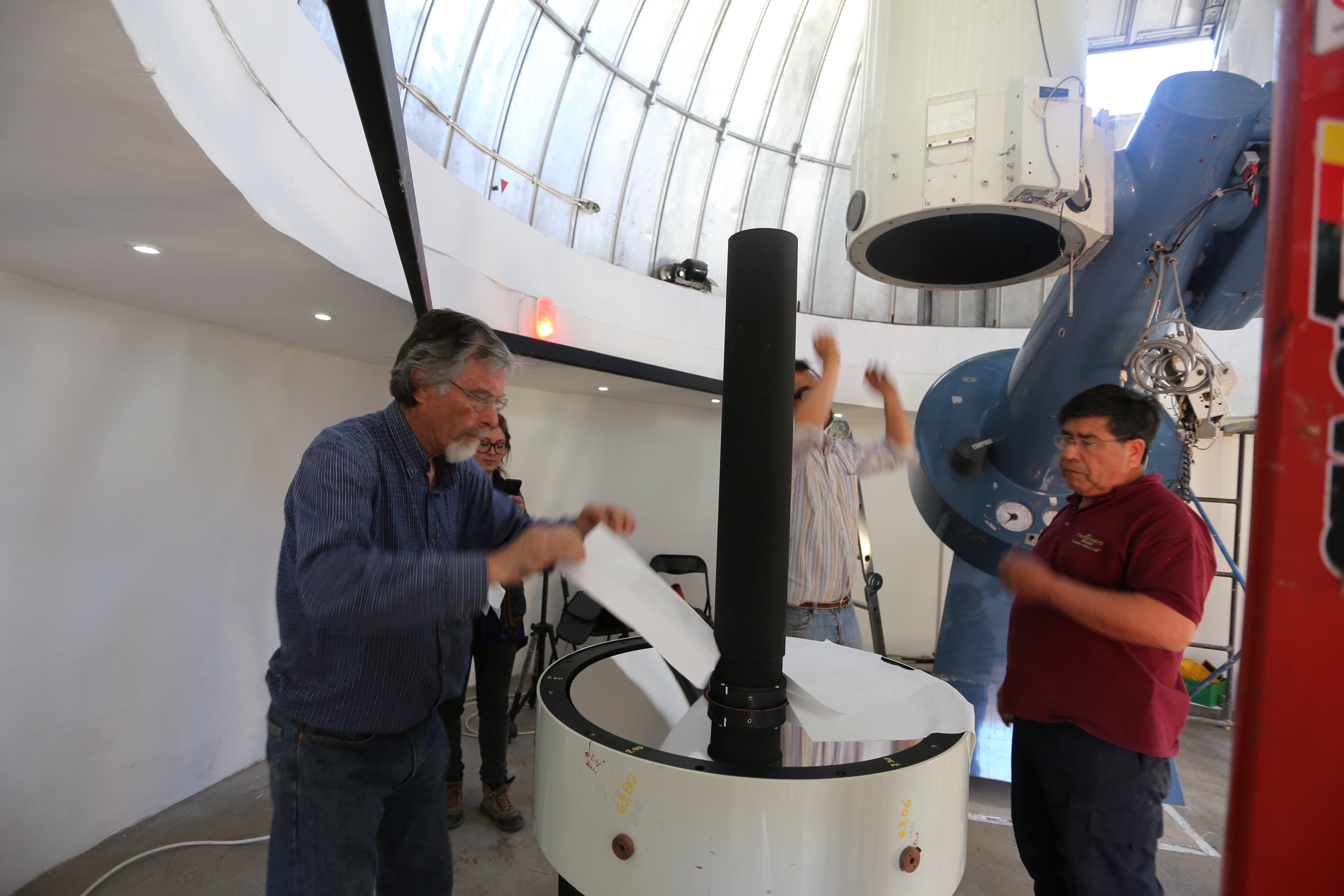}
	\caption{The technical support team working during the arrival of the primary (M1) and secondary (M2) mirror at the Observatorio Pocuro.}
	\label{fig:primary_mirror}
\end{figure}

Once the M1 and M2 mirrors were aluminized in La Silla Observatory they were delivered to the Observatorio Pocuro. The primary mirror was placed in the cell and the M2 was installed in the secondary ``spider'' support, finally both were properly collimated and aligned. 

\section{Revival Process: from analog electronic to web-based TCS}
\label{sec:revival}
The restoration process will give a second life to the abandoned Bochum 0.61-meter telescope which will be used now for science, teaching and outreach. This project will focused mainly on the upgrade of its Telescope Control System (TCS) and will be conducted by the company ObsTech SpA, based in Chile. The TCS has already been successfully deployed on La Silla 1-m telescope \cite{1mSilla}, bringing that telescope to modern standards. The TCS considers software running in a group of single-board computers interacting together as a network with the CoolObs TCS, potentially allowing the user to run observations remotely through a secured Internet web interface.

\subsection{Telescope status}
The overall status of the telescope is more than satisfactory. The telescope does not suffer from rust or deteriorated parts and also is well balanced. Painting is in good shape and the telescope can be operated using the current control system. 

The control system is running on an outdated and not operational DOS-based system. A weak aspect of the current settings is that spare parts for the electronics and encoders will be harder and harder to find, making future maintenance rather challenging. Tracking quality is good and smooth but could be improved if the control system could handle tracking using data from the telescope pointing model.

The right ascension (RA) and declination (DEC) axes are each driven by 2 asynchronous motors and 1 stepper motor. This configuration was used in order to be able to handle the high dynamical range of possible target speeds for the telescope (tracking, centering and slewing). Of course this configuration was driven by the hardware available at that time and could now be replaced by a simpler setup consisting of two DC motors (one for each axis) and their respective modern control and power electronics.

\begin{figure}[H]
	\centering
	\includegraphics[width=.7\textwidth]{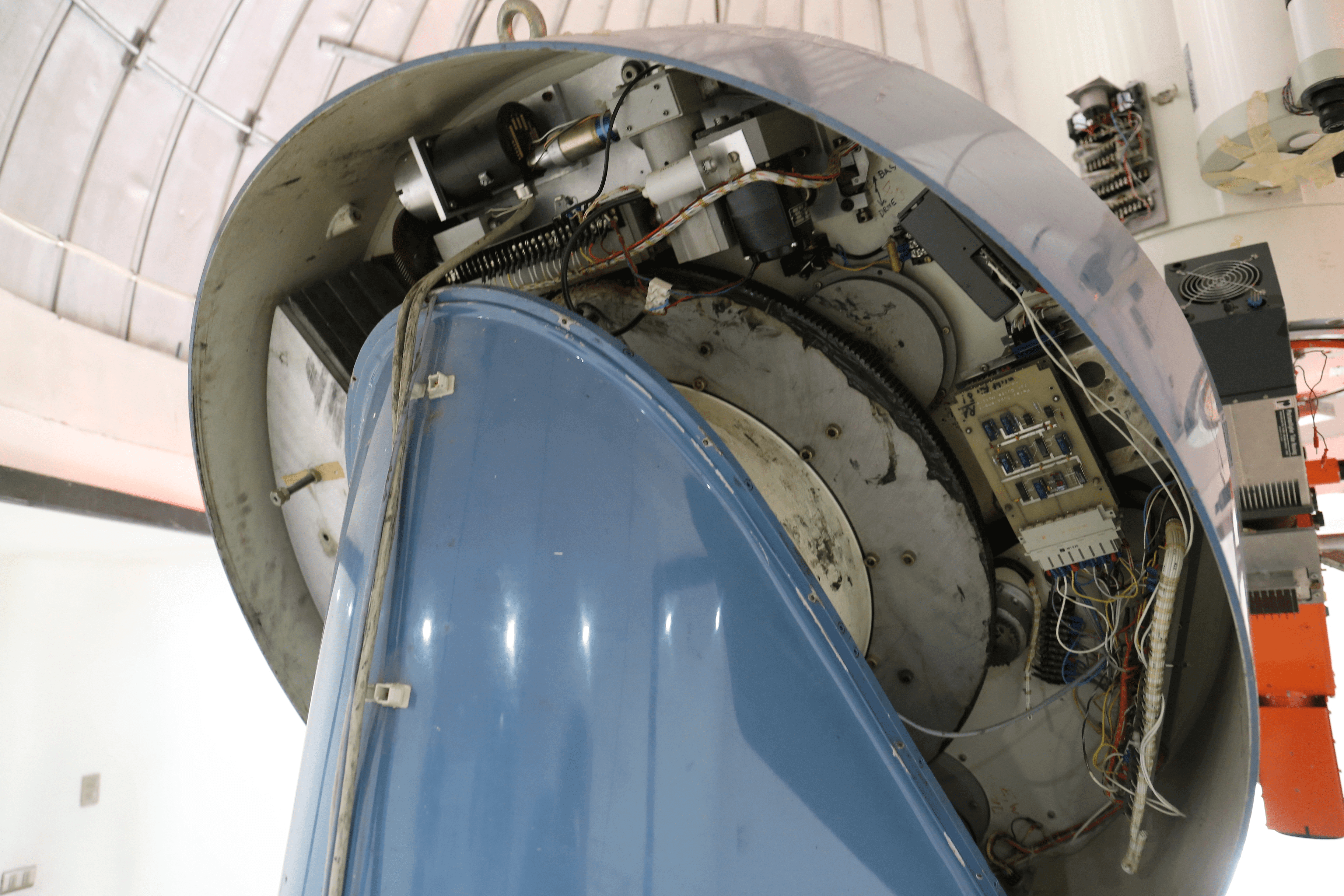}
	\caption{Right ascension motors and gears.}
	\label{fig:motor_gears}
\end{figure}

Current motors are 1/9 HP Bodine Electric AC motors giving approximately 65 oz-in torque and 1700 rpm max speed. At the beginning the idea is to replace them by 24V 1/8 HP DC brushed motors manufactured by the same company (N6424) or similar power Pittmann DC054B Series Brush DC motors which have the advantage of allowing a configuration that includes a 5000 ticks/revolution encoder on its axis. The actual decision whether to keep the existing stepper motor or to change them with DC motors previously mentioned will be taken at the next visit upon a thorough mechanical and electronical feasibility analysis.

In principle, the DC motor control feedback measurements would be: (1) encoders on the worm gear and (2) a high resolution encoder on the main axis of the telescope. In order to prevent and dynamically correct eventual periodic and high frequency tracking errors of the worm gears it will be necessary to replace the current axis encoder by a high resolution encoder on the main axis of the telescope. This will allow us to have a very high quality tracking even if the gears present large periodical errors or an eventual mechanical deterioration due to their extended use in the past decades. This upgrade is depending on further evaluation and mechanical installation feasibility which will be performed in the next on-site visit by ObsTech.

Despite the fact that the current encoders are of high quality and could in principle be reused, we consider it significantly safer to upgrade to more recent models because spares for the existing ones may only exist at La Silla’s component graveyard. It was proposed to use Gurley R158s encoders for the worm, giving an absolute precision of 15 arcsec with 25000 pulses/revolution and Gurley 8x60 encoder on the main axis offering a 0.25 arcsec relative precision and a 2 arcsec absolute positioning.

\begin{figure}[H]
	\centering
	\includegraphics[width=.7\textwidth]{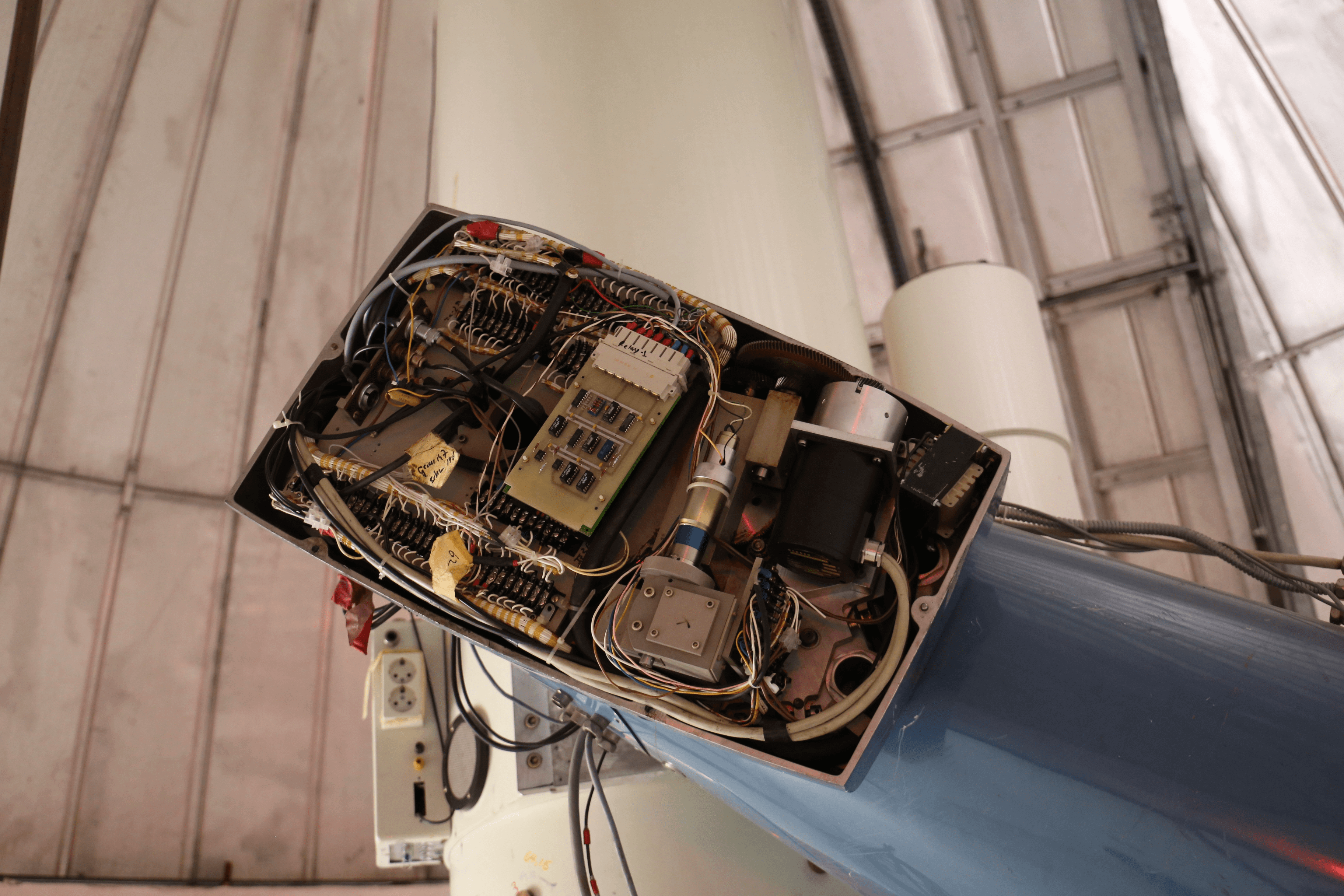}
	\caption{Current declination worm encoders.}
	\label{fig:worm_enconders}
\end{figure}

\begin{figure}[H]
	\centering
	\includegraphics[width=.7\textwidth]{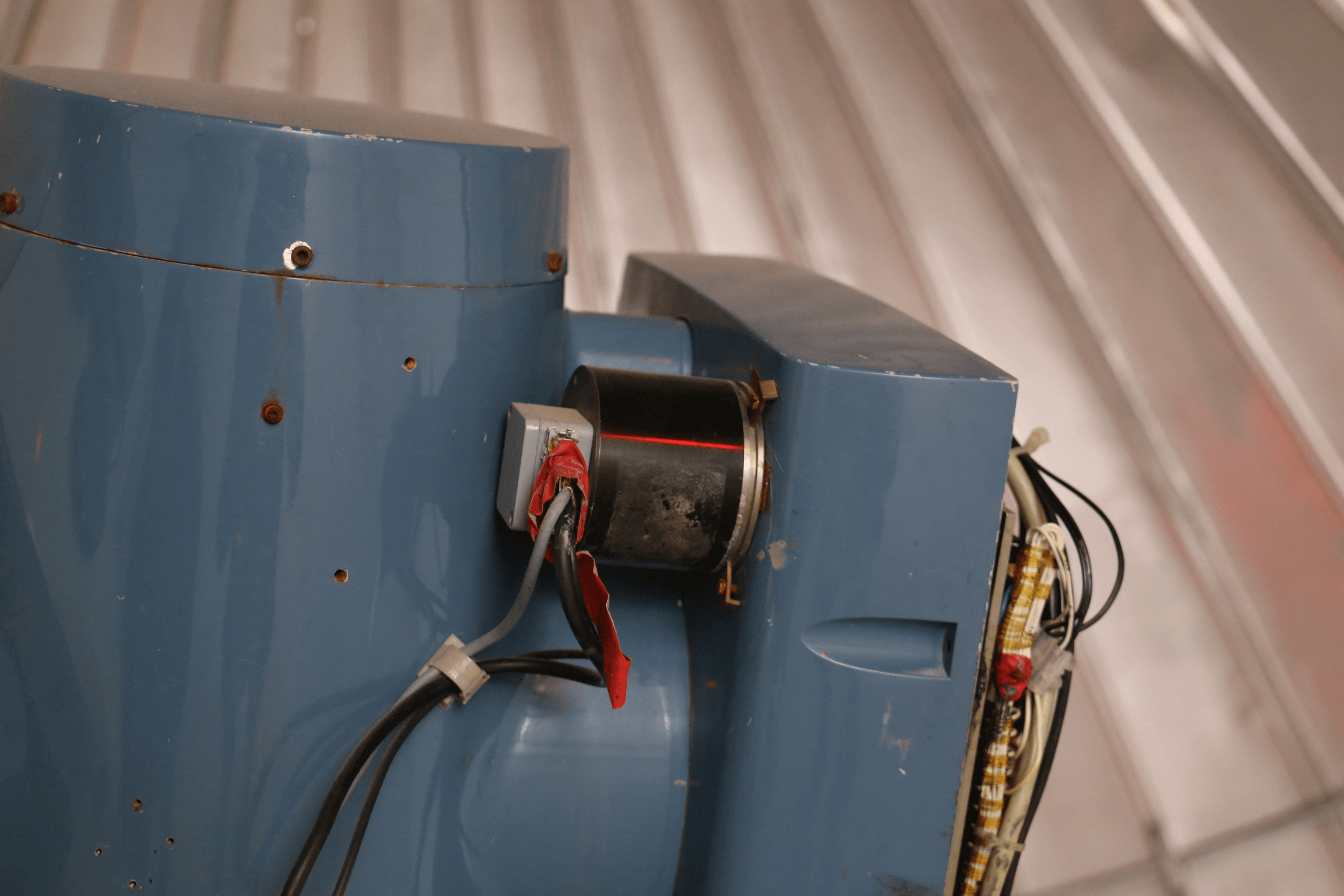}
	\caption{On-axis encoder which will be replaced.}
	\label{fig:onaxis_encoders}
\end{figure}

\subsection{Upgrade plan}
The integration of the telescope control system will be in four runs on site. The first one was done in early 2018 where the telescope evaluation was performed. The following run will be dedicated to a temporary installation and calibration of the new motors and encoders. The third run will aim for the final installation and testing. Finally, a fourth run will be used for final commissioning with on-sky observations and the generation of the pointing model. 
\newpage
The proposed upgrade plan consists of the following items which are presented with estimated timescales.

\begin{itemize}
\item Electronics, connectors and parts purchase and delivery to Santiago.
\item TCS box construction and lab testing: 2 months in Santiago. This involves soldering all components of the TCS into a custom made electronic box, and a thorough suite of lab tests.
\item TCS box temporary installation on site: During this second on-site run we will perform a temporary installation at the telescope, and calibrate the TCS with the electronic box and new hardware components connected to the telescope. Mechanical interfaces will be tested and refined on-site if possible or modifications slated for off-site machining. This run will involve mostly day-time work.
\item TCS box permanent installation at the telescope: During this third run the TCS will be permanently attached to the telescope and all systems will be tested. Mostly day-time work.
\item Final on-sky commissioning and pointing model generation: This is a commissioning run in which the system performance will be verified on-sky. One of the first tasks will be generation of the pointing model.
\end{itemize}

The telescope evaluation and electronics parts purchase was already done. Currently ObsTech SpA are working on the TCS box in their lab in Santiago and are performing preliminary tests. The next on-site run is expected for middle June 2018 when we will perform the temporary installation and calibration of the new hardware components to the telescope. 

\section{Conclusion}
The revival process of the Bochum 0.61 meter telescope, a former abandoned member of La Silla Observatory, was documented. This was a project with great challenges since its very inception which has nonetheless given us a valuable learning experience. Currently we are working on the renewal of the Bochum telescope in order to give give a ``Second life'' to this abandoned telescope and offer the students as well as the researchers of Universidad de Valparaíso a solid observational potential, as well as teaching and outreach opportunities.

One of the applications of the modernized Bochum telescope will be teaching both undergraduate and graduate students of the UV in the field of observational astronomy with the possibility to make observations on the site as well as remotely, and to familiarize with the procedures regarding telescope operations. Complementary to teaching our 0.6-m telescope will be used for research projects of the members of our Institute and will collaborate on several survey campaigns. 

Finally, but not less important, our Bochum telescope will have the strong outreach mission, we will offer 20\% of the time to the local community and the Observatorio Pocuro which will use it to bring the local community closer to the wonders of astronomical observation and science in general.

\acknowledgments
This project was suported by Universidad de Valparaiso through the project BPI 1619 - \textit{Instalación de un sistema de hardware y software de última tecnología en el telescopio UV en Pocuro, para observaciones científicas CCD presenciales y a control remoto}. Also we would like to thank the Ilustre Municipalidad de Calle larga for their commitment and collaboration. S. Zúñiga-Fernández acknowledge financial support from the ICM (Iniciativa Cient\'ifica Milenio) via the N\'ucleo Milenio de Formaci\'on Planetaria grant. 

\bibliography{report} 

\begin{thebibliography}{1}

\bibitem{hanuschik_1988}
Hanuschik, R.~W., ``Absolute spectrophotometry of {SN 1987A} and the
  {$H_{\alpha}$} {Bochum} event,'' {\em Publications of the Astronomical
  Society of Australia}~{\bf 7}(4),  446–449 (1988).

\bibitem{1mSilla}
Ropert, S., Suc, V., Jordán, A., Tala, M., Liedtke, P., and Royo, S., ``{TCS}
  and peripheral robotization and upgrade on the {ESO} 1-meter telescope at {La
  Silla Observatory},'' {\em Proc. SPIE}~{\bf 9912},  9912--9912--8 (2016).

\end{thebibliography}
\bibliographystyle{spiebib} 

\end{document}